%% file: main.tex
\documentclass{article}

\usepackage{iclr2027_conference,times}

\usepackage[utf8]{inputenc}
\usepackage[T1]{fontenc}
\usepackage{microtype}
\usepackage{amsmath,amssymb}
\usepackage{booktabs}
\usepackage{longtable}
\usepackage{makecell}
\usepackage{algorithm}
\usepackage{capt-of} 
\usepackage{algpseudocode}
\usepackage{float}
\usepackage{graphicx}
\usepackage{adjustbox} 
\usepackage[para]{footmisc} 
\usepackage{hyperref}
\usepackage{url}
\usepackage{xcolor}
\usepackage{soul}
\definecolor{hlorange}{HTML}{FFC680}
\definecolor{hlblue}{HTML}{9CD3FF}
\usepackage{colortbl} 
\usepackage{tabularx}
\definecolor{optdeep}{HTML}{4A2472}  
\definecolor{detdeep}{HTML}{1A3F7A}  
\usepackage{tikz}
\usetikzlibrary{arrows.meta,positioning,shapes.geometric}
\usepackage{pgfplots}
\pgfplotsset{compat=1.18}
\usepgfplotslibrary{groupplots}
\usepackage[skins,breakable]{tcolorbox}
\usepackage{enumitem}
\usepackage{listings}

\title{MetaVul: Agentic Harness Discovery for Vulnerability Detection}

\author{\adjustbox{max width=\textwidth}{Ira Ceka$^{1}$, Arnav Garg$^{1}$, Ryan Lutz$^{2}$, Ting Zhang$^{3}$, Ratnadira Widyasari$^{4}$, Gail Kaiser$^{5}$, Baishakhi Ray$^{1}$} \\[4pt]
\normalfont $^{1}$Columbia University \quad $^{2}$The Citadel \quad $^{3}$Monash University \\
\normalfont $^{4}$University of Victoria \quad $^{5}$Columbia University, Department of Computer Science \\[2pt]
\normalfont \texttt{iceka@cs.columbia.edu} \quad \texttt{ag5516@columbia.edu} \quad \texttt{rlutz@student.citadel.edu} \\
\normalfont \texttt{Ting.Zhang@monash.edu} \quad \texttt{ratnadiraw@uvic.ca} \\
\normalfont \texttt{kaiser@cs.columbia.edu} \quad \texttt{rayb@cs.columbia.edu}
}

\input{macros}
\iclrfinalcopy 

\begin{document}

\maketitle
\lhead{Preprint.} 

\begin{abstract}

Large Language Model (LLM) agents are increasingly used to automate vulnerability detection, but reliable detection is challenging because agents must distinguish subtle, security-relevant fixes from benign code variation. While underlying LLM capability has improved significantly, the accuracy of the agent still depends heavily on the \emph{harness}: the code that defines the scaffolding around the LLM. Harnesses are typically fixed, and largely hand-written, limiting adaptation to evaluator feedback. Additionally, when harnesses are static, improvements on one dataset or model may not necessarily transfer to others. To address these challenges, we introduce MetaVul, a self-improving framework for harness discovery in vulnerability detection. Starting from a seed structured reasoning scaffold, MetaVul uses a workflow-based optimization loop to iteratively refine the harness using feedback from paired vulnerable and patched functions, all while preserving the structured reasoning and verdict format. At each iteration, it searches over the full history of prior candidate harnesses, with access to their trajectories, logs, and metrics, and it returns a set of Pareto-optimal harnesses rather than a single best candidate. We apply MetaVul to discover agent harnesses for architectures including Codex CLI and Claude Code. A validation set (202 pairs) from TitanVul and CWEval is used for optimization, while the unseen PrimeVul dataset (430 pairs) is kept as the held-out set. We later test whether harnesses learned from one model transfer to others not seen during optimization. Results on both frontier and small LLMs show that the optimized harnesses improve pairwise vulnerability detection up to 23.3 percentage points on the validation set (Opus 4.6) and up to 25.8 points on the held-out set with a Pareto-frontier harness (DeepSeek V4 Flash). Harnesses learned with one model also transfer to others. We show that harnesses optimized with Opus 4.6 transfer to Opus 5, improving held-out pairwise accuracy by 14.7 points without the need for further optimization.
Code is available at \url{https://github.com/ARiSE-Lab/MetaVul}~\citep{metavul_repo}.
\end{abstract}

\section{Introduction}
\label{sec:intro}


Artificial intelligence (AI) agents utilize a large language model (LLM) as their core reasoning component for taking actions toward a given goal. “Vibe coding,” or using AI tools, particularly by way of coding agents, is becoming increasingly prevalent across the software industry. Consequently, the growing volume and pace of AI-assisted code generation increase the importance of effective vulnerability detection. LLM agents rely not only on the capabilities of their underlying model, but also on the harnesses that structure how those agents reason, interact with tools, and arrive at a decision. Designing these harnesses, known as \emph{harness engineering}, has become a practice of its own~\citep{openai2026harness, anthropic2026harness}. For vulnerability detection, these harnesses are typically hand-designed and are fixed during deployment, even though their structure and context can substantially alter an agent’s results. In practice, vulnerability detection is substantially affected by how the agent inspects the code, and how the harness will shape that workflow~\citep{yildiz2025benchmarkingllmsllmbasedagents}. This understanding directly motivates a harness that can learn its design systematically through evaluation feedback.

Recent work studies the evaluation, evolution, and adaptation of agent harnesses using graded scores, verifiers, execution traces, or prior experience~\citep{lee2026metaharnessendtoendoptimizationmodel,ursekar2026harnessoptbenchevaluatingllmsharness,huang2026evobenchlanguagemodelsimprove,zhang2026darwinxevolvingagentharnesses,huang2026memoharnessagentharnesseslearn,esakkiraja2026starharnessevolvingharnessesstratified,nie2026tthetesttimeharnessevolution}, while OneDayAgent develops a fixed long-horizon harness~\citep{zheng2026onedayagentlonghorizonharnessautonomous}; however, none investigates harness optimization for vulnerability detection or incorporates security-specific objectives such as CWE identification, vulnerability localization, and vulnerable-patched discrimination.
Harness engineering for vulnerability detection, however, poses challenges.

Firstly, \textbf{false alarms are costly.} In practical security settings, analysts receive far more alerts than they can effectively triage. Several industry studies identify alert fatigue and analyst burnout as major problems~\citep{tariq2025alertfatigue}. This problem is well documented among traditional vulnerability-detection tools~\citep{charoenwet2024empiricalstudystaticanalysis,sajadi2026empiricalanalysiscodeqlfalse} and persists with recent LLM-based detectors~\citep{li2026llmbasedvulnerabilitydetectionproject}. As recently as April 2026, an independent security company reported poor discriminative performance on PrimeVul~\citep{primevul}\footnote{Following PrimeVul, we adopt \emph{pairwise accuracy} as a core metric alongside accuracy and F1. High pairwise accuracy indicates that the detector can distinguish vulnerable code from its fixed version, reducing false alerts and, consequently, alert fatigue. We explain this further in Sec.~\ref{sec:metrics}.}; even Claude Opus 4.6, a flagship model shown by Anthropic to discover previously unknown vulnerabilities~\cite{anthropic2026zerodays}, struggled to reliably distinguish vulnerable functions from their patched counterparts~\citep{zeropath2026opus46}. These findings further our motivation and demonstrate the outstanding and relevant nature of vulnerability discrimination, which remains an open and significant problem in practice. Moreover, these findings reinforce the notion that strengthening the underlying models does not eliminate the need for a robust design surrounding the agentic workflow. Rather, harnesses and their associated design choices matter as they directly affect downstream behavior.

Secondly, \mbox{\textbf{harness optimization is non-trivial.}} A vulnerability detection harness controls how an agent inspects code, how it structures its analysis, and how it reasons about security. Vulnerability detection as a task, however, poses a \emph{multi-objective} challenge. A strong harness not only has to accurately classify vulnerabilities, but also localize them, identify their CWE(s), ground the conclusions in code evidence, as well as distinguish the vulnerable functions from their patched counterparts. \mbox{\citet{primevul}} show why traditional classification metrics, such as accuracy and F1, are insufficient on their own: models that perform well on conventional benchmarks approach random guessing on PrimeVul's paired evaluation. Optimizing for a single metric may therefore reward superficial patterns rather than robust vulnerability reasoning. HarnessDev~\mbox{\citep{wu2026harnessdevllmscreateevolve}} shows that a harness can overfit to the model that wrote it, memorize development examples, or improve one capability while silently regressing another. It can also overfit its search tasks: \mbox{\citet{wang2026rethinkingevaluationharnessevolution}} found that, when harness evolution and evaluation share the same tasks, the resulting improvements largely disappear on held-out tasks, which suggests that optimization can learn task-specific shortcuts rather than generalizable harness improvements.

Finally, \mbox{\textbf{defense must keep pace with AI-driven offense.}} With the rise of AI-driven exploit generation, attackers are moving at immense speed. In October 2026 alone, two critical PHP vulnerabilities were disclosed after AI-assisted research revealed that a 14-year-old memory-safety flaw in the PHP runtime engine allowed unauthenticated attackers to execute malicious code remotely on servers hosting PHP-based web applications, triggering a global security crisis~\citep{php2shell}. These attacks involve auditing millions of lines of code, mapping out reliable exploit strategies, and adapting to multiple configurations. While offensive security is bottlenecked by compute, defense has another challenge to tackle: to survive this offensive scale, defenders must automate triage and blocking, but doing so can be catastrophic due to the false positives mentioned previously. It is thus paramount for defense to have scalable methods for identifying suspicious code and evaluating, triaging, and applying potential fixes quickly.

We address these challenges with \mbox{\textbf{MetaVul}}, which improves vulnerability detection through iterative harness construction, driven by validation feedback. The ``\textbf{meta}'' in MetaVul comes from the way MetaVul finds better ways to find vulnerabilities, with the optimizer and detector operating on two different levels: the optimizer designs the harness, and the detector uses it. A sandboxed detection agent evaluates each candidate harness during the discovery phase, and MetaVul returns a Pareto frontier of harnesses that trade off discrimination and localization ability. The discovered harnesses are then evaluated against held-out datasets and unseen models, hence testing generalization.
We ask whether MetaVul improves vulnerability-detection harnesses and whether the gains transfer to unseen data and unseen models (\mbox{\textbf{RQ1}}). We then evaluate how MetaVul compares with other LLM-based or multi-agent vulnerability detection methods (\mbox{\textbf{RQ2}}). Finally, we ask whether MetaVul's gains rely on ``shortcut'' exploitation, like data leakage signals, embedded answers, or trivial verdict bias (\mbox{\textbf{RQ3}}).
We find that MetaVul improves pairwise accuracy by 11.4 to 23.3 points (\mbox{$1.6\times$}-\mbox{$4.3\times$}) on validation and by up to 25.8 points (\mbox{$2.9\times$}) on 430 unseen PrimeVul pairs. Harnesses also transfer to models that took no part in optimization, with gains up to 14.7 points (\mbox{$1.7\times$}). MetaVul outperforms prior LLM-based and multi-agent detectors and does so without evidence of ``shortcut'' exploitation.


\section{Related Work}
\label{sec:related-work}

Previous work has explored LLM-based vulnerability detection, harness engineering and optimization, and structured reasoning strategies for code.
\textbf{LLM-based vulnerability detection.}
\citet{primevul} show that LLMs struggle to distinguish vulnerable functions from their patched versions under paired evaluation. They propose pairwise accuracy as a stricter alternative to conventional metrics, and also introduce PrimeVul, which addresses the noisy labels and train-test duplication of earlier datasets through stricter labeling, de-duplication, and a chronological split. Chain-of-thought (CoT) prompting~\citep{wei2022chain} improves pairwise performance over two-shot prompting but stays below random guessing.
GPTLens~\citep{gptlens} pairs an auditor that proposes candidate vulnerabilities with a critic that filters false positives, VulTrial~\citep{vultrial} makes use of courtroom-style agents that debate a verdict, and ConfColl~\citep{confcoll} escalates from a single agent to retrieval-augmented and multi-agent reasoning when needed.
All four methods use GPT-4o as the backbone~\citep{gpt4o,vultriage}, and we compare against them in Sec.~\ref{sec:rq3}. Unlike GPTLens, VulTrial and ConfColl, MetaVul keeps a single detector agent and instead optimizes its harness.
At project scale, \citet{li2026llmbasedvulnerabilitydetectionproject} report low recall, high false-discovery rates and high computational cost.
\textbf{Harness engineering and optimization.}
Recent work treats the agent harness (prompts, tools, memory, context management and control flow) as an optimization target independent of model weights~\citep{lee2026metaharnessendtoendoptimizationmodel,ursekar2026harnessoptbenchevaluatingllmsharness,openai2026harness,anthropic2026harness}, in contrast to weight-level methods such as ORPO~\citep{hong2024orpomonolithicpreferenceoptimization}.
GEPA~\citep{agrawal2026gepareflectivepromptevolution} optimizes prompts from execution trajectories and model feedback, and SecureForge~\citep{liu2026secureforgefindingpreventingvulnerabilities} optimizes system prompts for secure code generation using automated vulnerability auditing.
Meta-Harness, MemoHarness, DarwinX, StarHarness, and TTHE explore complementary forms of harness adaptation using execution experience, search, held-out selection, and test-time adaptation~\citep{lee2026metaharnessendtoendoptimizationmodel,huang2026memoharnessagentharnesseslearn,zhang2026darwinxevolvingagentharnesses,esakkiraja2026starharnessevolvingharnessesstratified,nie2026tthetesttimeharnessevolution}, and OneDayAgent supports long-horizon execution with task decomposition, persistent memory and output verification~\citep{zheng2026onedayagentlonghorizonharnessautonomous}.
HarnessOpt-Bench, Evo-Bench and HarnessDev evaluate whether models can build or improve harnesses that generalize beyond their optimization tasks~\citep{ursekar2026harnessoptbenchevaluatingllmsharness,huang2026evobenchlanguagemodelsimprove,wu2026harnessdevllmscreateevolve}, and \mbox{\citet{wang2026rethinkingevaluationharnessevolution}} underscore the need for held-out evaluation, showing that any apparent gains might vanish on held-out tasks, or simply reflect that extra compute was spent searching over candidates.
\textbf{Reasoning strategies for code.}
Structured chain-of-thought (SCoT) organizes reasoning around program structures~\citep{li2023structured}, Modularization-of-Thought (MoT) decomposes it into hierarchical modules~\citep{pan2025mot}, and Tree of Thoughts (ToT) explores multiple reasoning paths~\citep{yao2023tree}.
Despite this, to our knowledge, none of these methods have explored harness optimization for vulnerability detection, nor tested whether such harnesses transfer to unseen datasets and models. MetaVul automatically optimizes harnesses for this task, selects them on security-specific objectives (telling vulnerable functions apart from their fixes, i.e., pairwise accuracy, and joint CWE and line localization), and evaluates them on an unseen dataset and on models that took no part in optimization.

\section{Methodology}
\label{sec:methodology}

\input{sections/fig_pipeline}

MetaVul separates harness discovery from evaluation on unseen data and models. Beginning with a seed harness (\mbox{$h_0$}), the optimizer (\mbox{$\mathcal{A}_{\mathrm{opt}}$}) proposes a new harness, using \emph{validation feedback}. The isolated detector (\mbox{$\mathcal{A}_{\mathrm{det}}$}) evaluates each harness candidate. After discovery, MetaVul retains the Pareto-optimal harnesses and evaluates them on held-out PrimeVul using the search and transfer models. Fig.~\mbox{\ref{fig:pipeline}} presents the full pipeline, and Algorithm~\mbox{\ref{alg:metavul}} details the discovery loop.

\begin{figure}[t]
\centering
\begin{minipage}[t]{0.58\linewidth}
\vspace{0pt}
\fontsize{7.8}{9.4}\selectfont
\algrenewcommand{\algorithmiccomment}[1]{\hfill{\fontsize{6.5}{8}\selectfont\color{black!65}$\triangleright$~#1}\hspace*{12pt}}
\noindent\rule{\linewidth}{0.8pt}\par\vspace{1pt}
{\small\refstepcounter{algorithm}\label{alg:metavul}\noindent\textbf{Algorithm~\thealgorithm}\enspace\textsc{MetaVul} harness discovery loop\par}
\vspace{1pt}\noindent\rule{\linewidth}{0.4pt}\par
\vspace{-2pt}
\begin{algorithmic}[1]
\Require Seed harness $h_0$, validation pairs $D$, maximum iterations $T$, patience $K$, optimizer agent $\mathcal{A}_{\mathrm{opt}}$, detection agent $\mathcal{A}_{\mathrm{det}}$
\State $(m_0, z_0) \gets \textsc{Evaluate}(\mathcal{A}_{\mathrm{det}}, h_0, D)$ \Comment{$m$: metrics; $z$: predictions}
\State $\mathcal{H} \gets \{(h_0,m_0,z_0)\}$; $h_{\mathrm{best}} \gets h_0$
\State $s_{\mathrm{best}} \gets \textsc{PairwiseAcc}(z_0)$; $c \gets 0$
\For{$t \gets 1$ to $T$}
    \State $W_t \gets \textsc{BuildWorkspace}(\mathcal{H}, h_{t-1})$
    \State $h_t \gets \textsc{Propose}(\mathcal{A}_{\mathrm{opt}}, W_t)$ \Comment{keeps SCoT and verdict schema}
    \If{$h_t$ is invalid or the optimizer fails}
        \State $c \gets c+1$; \textbf{continue}
    \EndIf
    \State $(m_t, z_t) \gets \textsc{Evaluate}(\mathcal{A}_{\mathrm{det}}, h_t, D)$ \Comment{fresh call per function}
    \State $s_t \gets \textsc{PairwiseAcc}(z_t)$
    \State $\mathcal{H} \gets \mathcal{H} \cup \{(h_t,m_t,z_t)\}$
    \If{$s_t > s_{\mathrm{best}}$}
        \State $h_{\mathrm{best}} \gets h_t$; $s_{\mathrm{best}} \gets s_t$; $c \gets 0$
    \Else
        \State $c \gets c+1$
    \EndIf
    \If{$c \ge K$} \textbf{break} \EndIf
\EndFor
\State $\mathcal{F} \gets \textsc{ParetoFront}(\mathcal{H};\ \textsc{PairwiseAcc}, \textsc{Joint})$ \Comment{non-dominated}
\State \Return $\mathcal{F}$
\end{algorithmic}
\vspace{-4pt}\noindent\rule{\linewidth}{0.4pt}
\end{minipage}\hfill
\begin{minipage}[t]{0.39\linewidth}
\vspace{0pt}
\centering
\small
\captionof{table}{Datasets used by MetaVul. A circled number corresponds to its respective pipeline stage in Fig.~\ref{fig:pipeline}.}
\label{tab:datasets}
\setlength{\tabcolsep}{4pt}
\renewcommand{\arraystretch}{1.1}
\begin{tabularx}{\linewidth}{lX}
\toprule
\multicolumn{2}{l}{\textbf{Validation set $D$}} \\
\midrule
\rowcolor{gray!8}Dataset   & TitanVul + CWEval \\
Pairs     & 202 \\
\rowcolor{gray!8}Languages & 161 C, 41 C++ \\
CWEs      & 72 \\
\rowcolor{gray!8}Used for  & \scalebox{0.8}{\stepnum{1}} Validation \\
\addlinespace[3pt]
\bottomrule
\end{tabularx}

\vspace{6pt}
\begin{tabularx}{\linewidth}{lX}
\toprule
\multicolumn{2}{l}{\textbf{Held-out set}} \\
\midrule
\rowcolor{gray!8}Dataset   & PrimeVul \\
Pairs     & 430 \\
\rowcolor{gray!8}Languages & C/C++ \\
CWEs      & 61 \\
\rowcolor{gray!8}Used for  & \scalebox{0.8}{\stepnum{2}} Held-out evaluation \\
          & \scalebox{0.8}{\stepnum{3}} Cross-model transfer \\
\addlinespace[3pt]
\bottomrule
\end{tabularx}
\end{minipage}
\end{figure}

\textbf{Harness discovery and selection.} MetaVul iteratively generates and evaluates candidate harnesses before selecting those that best balance its evaluation objectives.
\textbf{\textit{Discovery loop.}} MetaVul first evaluates the seed harness $h_0$. At each iteration $t$, the optimizer uses previous harnesses and results to propose $h_t$, which the detector evaluates on $D$. The results are added to the search history. Discovery ends after $T$ iterations or $K$ iterations without improving pairwise accuracy.
\textbf{\textit{Agent roles.}} The optimizer \mbox{$\mathcal{A}_{\mathrm{opt}}$} revises the harness using validation feedback. The detector \mbox{$\mathcal{A}_{\mathrm{det}}$} runs the proposed harness and returns predictions for scoring. The agents may use the same model, but they run separately and never share state.
\textbf{\textit{Pareto selection.}} After the loop, MetaVul returns the Pareto frontier of all evaluated harnesses over two validation objectives, pairwise accuracy and joint hit rate (Sec.~\ref{sec:metrics}), both maximized (\textsc{ParetoFront} in Algorithm~\ref{alg:metavul}). \supp{S6_additional_results}{S6} gives a worked example.
We say harness A \emph{dominates} harness B if A scores no lower than B on both pairwise accuracy and joint hit rate, and higher than B on at least one of those metrics. The frontier \mbox{$\mathcal{F}$} contains those harnesses which are not dominated by any other harness. If two or more harnesses have identical scores on both metrics, we keep only the earliest one.

\label{sec:sandboxing}\textbf{Sandboxing.} It is essential that the held-out set is never exposed to the agent during discovery (the ``optimization'' phase); otherwise, contamination would make any held-out gains indistinguishable from memorization. Neither discovery nor evaluation should be influenced by system files, default instruction files (such as \texttt{CLAUDE.md}) or user settings. To address this, we sandbox every agent. The optimizer and detector agents each run isolated CLI processes that can neither request nor receive any additional permissions. Each agent is limited in the data that it can access. As shown in Fig.~\ref{fig:pipeline}, the agent and its model serve two roles: an optimizer and a detector (which we also call the evaluator).
\textbf{\textit{Optimizer.}} The optimizer can only access its workspace, which holds the validation history: earlier harnesses, their metrics, and their predictions with labels. After each iteration, its results are added to the workspace of the next iteration. \supp{S10_context_budget}{S10} describes the context budget. For Claude Code, the working directory is set to the workspace, the read tool is restricted to it, write and edit tools are denied, and shell access is limited to read-only commands. No \texttt{CLAUDE.md} files, settings, or memory related to the user or project are loaded into the context. For Codex, an OS-level sandbox grants read-only access to the workspace, with no write or network access.
\textbf{\textit{Detector.}} The detector is launched freshly for every function. To ensure a non-contaminated run, it has no workspace access (i.e., from the optimizer), no labels, no trajectory, and no memory between calls. The detector receives only the rendered harness and one function. For this phase, Claude Code runs from a \texttt{/tmp} directory with no tools and an empty system prompt. Similarly, Codex runs the harness in an empty temporary directory tree with read-only access and no network connection.
The predictions from the held-out PrimeVul evaluation and the cross-model transfer are never added to any workspace (only the harness changes between iterations). Detailed agent restrictions are given in \mbox{\supp{S8_agent_sandboxing}{S8}}.

\section{Experimental Design}

%

We outline our experimental setup: the datasets, seed harness, agents, models, and configurations (Sec.~\ref{sec:datasets}), and the evaluation metrics (Sec.~\ref{sec:metrics}).

\subsection{Datasets, Model, and Agent Selection}
\label{sec:datasets}

\textbf{Datasets.} We use separate datasets for optimization and testing, so that we can measure how harness improvements carry over to an unseen dataset and to models that took no part in optimization.
Every sample is a vulnerable function paired with its patched version.
Table~\ref{tab:datasets} summarizes both splits.

\begin{itemize}[leftmargin=1.2em, itemsep=2pt, topsep=2pt]
    \item \textbf{Validation set $D$.}\label{sec:validation-set} We handcraft a dataset $D$ from TitanVul and CWEval. TitanVul~\mbox{\citep{titanvul}} combines seven public vulnerability datasets and removes duplicate samples. Multiple LLM agents check fixes, keeping those that address genuine vulnerabilities understandable from the function alone. These checks reduce incorrect labels and unrelated code changes. We select 150 C/C++ pairs, sampled by CWE in proportion to the CWE's frequency (\supp{S1_validation_set}{S1}). We enforce some strict criteria, such as ensuring each CWE receives a proportional share to its TitanVul occurrence, there is a minimum of one of every CWE, even if rare, and pairs with unlabeled CWEs are also considered.
    CWEval~\citep{cweval} evaluates both the functional correctness and the security of LLM-generated code; each of its security-critical tasks has functional and security tests and a secure and an insecure reference implementation.
    We use all 52 of its C and C++ tasks, pairing each insecure implementation with its secure one.
    Together these form \mbox{$D$}: 202 pairs (404 functions).
    Because 79 of the 150 TitanVul pairs have no CWE label, CWE accuracy on \mbox{$D$} is computed over the remaining pairs which do.
    \item \label{sec:primevul}\mbox{\textbf{Held-out set.}} PrimeVul~\mbox{\citep{primevul}} addresses several issues of earlier datasets, including noisy labels and duplicate functions. PrimeVul contains vulnerable functions and their fixes from real open-source C/C++ projects, and we use its paired test set as our held-out ``unseen'' set, as this is a hard benchmark on which LLMs reach low pairwise accuracy.
    Because TitanVul aggregates public vulnerability datasets, we checked it for overlap with \mbox{$D$}: 5 PrimeVul pairs contain the same functions and fixes as TitanVul pairs in \mbox{$D$}. We remove them, leaving 430 pairs (860 functions) from 141 projects.
    Each vulnerable function in PrimeVul has a CWE label. The optimizer never sees PrimeVul data.
\end{itemize}

\mbox{\textbf{Seed harness.}} The harness is a standalone Python file that prepares a function for the detection agent and parses the agent's verdict. Every run begins with the same 151-line SCoT seed harness~\mbox{\citep{li2023structured}}, provided in \mbox{\supp{S3_seed_harness}{S3}}.\footnote{In \href{https://github.com/ARiSE-Lab/MetaVul/blob/main/results/prelim-scot.md}{preliminary experiments}, SCoT achieved higher accuracy and precision than a plain instruction, MoT, and ToT. Its structure also supports reasoning about sequences, branches, loops, and patch effects.}

\mbox{\textbf{Agent selection.}} We use Codex CLI and Claude Code. In each discovery session, the optimizer \mbox{$\mathcal{A}_{\mathrm{opt}}$} and detector \mbox{$\mathcal{A}_{\mathrm{det}}$} run the same CLI agent and model (only their access differs). The optimizer can read earlier harnesses, metrics, and labeled predictions in its workspace. The detector sees only the harness and one function (Sec.~\mbox{\ref{sec:sandboxing}}). For unseen-dataset and cross-model transfer, only the detector (\mbox{$\mathcal{A}_{\mathrm{det}}$}) runs, using the frozen harnesses to evaluate PrimeVul.

\textbf{Model selection.} We optimized harnesses with Inkling-Small~\citep{tinker_models}, DeepSeek V4 Flash~\citep{openrouter_deepseek}, Opus 4.6 and GLM-5.3~\citep{openrouter_glm}. We tested cross-model transfer on Opus 5, Nemotron 3.5 Lightning~\citep{tinker_models,openrouter_nemotron}, GPT-4o~\citep{openrouter_gpt4o}, GLM-4.7 Flash~\citep{openrouter_glm47} and GPT-OSS-120B~\citep{openrouter_gptoss}. 
These models cover smaller and frontier models, including open-weight and closed models. Opus 4.6 ran through our Claude subscription, whose usage limits every optimizer and detector call draws on, so we chose the remaining models from per-token providers (OpenRouter and Tinker; \supp{S12_inference_prices}{S12}).

\textbf{Configurations.} Opus 4.6 and Opus 5 ran through our Claude Code subscription, while all other models were applied through a local LiteLLM proxy to our external model providers (OpenRouter or Tinker). All runs allow up to 15 refinement iterations. Inkling-Small and DeepSeek V4 Flash ran all 15; GLM-5.3 used a patience of 4 and stopped after iteration 8; Opus 4.6 ran four, limited by subscription usage. Additional serving, timeout, and decoding configurations are outlined in \supp{S9_run_configurations}{S9}.

\subsection{Metrics}
\label{sec:metrics}

We evaluate each harness on its detection, discrimination, and localization ability.
\textbf{\textit{Traditional detection metrics.}} We treat each function as a binary decision with vulnerable as the positive class, and report accuracy, precision, recall and F1 over all functions.
\textbf{\textit{Pairwise accuracy.}} Let pair $i$ contain a vulnerable function $x_i^{v}$ and its patched version $x_i^{p}$. For a binary detector $f$, pairwise accuracy over $N$ pairs is $\operatorname{PairAcc}(f) = \tfrac{1}{N}\sum\nolimits_{i=1}^{N} \mathbf{1}[f(x_i^{v})=1 \land f(x_i^{p})=0]$.
A pair counts only when both decisions are correct, which makes the metric stricter than ordinary accuracy.\footnote{A detector that labels both the vulnerable function and its patch as vulnerable gets 50\% accuracy and full recall, but $\mathrm{pAcc}=0$. Hence, pAcc evaluates the extent to which the detector is able to identify the fix. \supp{S4_pairwise_accuracy_example}{S4} gives a worked example with three detectors.}
Pairwise accuracy is the target metric used in Algorithm~\ref{alg:metavul}.
\textbf{\textit{CWE and localization metrics.}} These score only true positives, the vulnerable functions $i$ that the harness flags, and only those with the relevant ground truth: a set of CWEs $C_i$ (usually one, sometimes several) and the gold lines $G_i$.\label{sec:gold-lines} None of the three datasets marks which lines are vulnerable, so we derive $G_i$ from each pair: it contains the lines of the vulnerable function that are modified or deleted by the fix, or the lines around the insertion point if the fix only adds lines.\footnote{The datasets pair vulnerable code with a benign associated counterpart. We make use of the lines changed by the fix as the closest approximation of the vulnerability position.} For CWEval, the fix is the secure reference implementation. For each such function, the harness returns an ordered list of CWE guesses, whose first guess is $\hat{c}_i^{(1)}$, and a set of evidence lines $\hat{G}_i$. \textit{CWE accuracy} is the percentage whose CWE list contains a CWE in $C_i$. \textit{Loc F1} is the harmonic mean of line-level precision $|\hat{G}_i \cap G_i| / |\hat{G}_i|$ and recall $|\hat{G}_i \cap G_i| / |G_i|$, each averaged over functions, and \textit{loc hit rate} is the percentage with at least one correct line ($\hat{G}_i \cap G_i \neq \emptyset$).
\textbf{\textit{Joint hit rate.}} Joint (\textsc{Joint} in Algorithm~\ref{alg:metavul}) asks whether a correctly flagged function is also explained correctly: the right vulnerability class in the right place. Let $T$ be the set of flagged vulnerable functions that have both $C_i$ and $G_i$. Function $i$ counts as correct only when its first CWE guess is in $C_i$ and it has at least one correct line: $\operatorname{Joint} = \tfrac{1}{|T|}\sum\nolimits_{i \in T} \mathbf{1}[\hat{c}_i^{(1)} \in C_i] \cdot \mathbf{1}[\hat{G}_i \cap G_i \neq \emptyset]$.
Only the first CWE guess is scored, so a harness cannot earn credit by listing several candidate CWEs.
Regarding parse failures, a function without a parseable verdict is counted as incorrect and stays in the denominator for all metrics, so every PrimeVul result covers all pairs.\footnote{Errors from a parse failure are rare, occurring at 0.36\% throughout and at most 0.63\% on any individual run.}

\input{results/pairwise_evolution_data}

\section{Discovery and Transfer}
\label{sec:results}
\label{sec:rq1}
\label{sec:rq2}

\input{results/pareto_scatter}

\input{results/pairwise_evolution_plot}

\textbf{Motivation \& Approach.} We ask whether a feedback-driven design can improve vulnerability-detection harnesses over a starter seed. We consider metrics such as pairwise accuracy and localization accuracy, as a strong harness should both tell vulnerable code apart from its fix and be able to locate the flaw. However, as mentioned, a harness must balance multiple objectives at once, and a gain on one metric may result in degradation on another, affecting how the best harness is selected. We therefore explore whether our approach improves pairwise accuracy and how these improvements affect other metrics, such as localization. Gains attained on the data used for optimization could still reflect overfitting: the harness search could learn task-specific shortcuts that work on the \emph{validation set} but disappear on the held-out tasks, or the harness can overfit to the model that wrote it, performing well on the model used for optimization but not transferring to other models~\citep{wang2026rethinkingevaluationharnessevolution,wu2026harnessdevllmscreateevolve}. This motivates \textbf{RQ1:} \textbf{\emph{Does feedback-driven search improve vulnerability-detection harnesses, and do the gains transfer to unseen data and models?}} We conduct four optimization runs using Codex CLI and Claude Code with Inkling-Small, DeepSeek V4 Flash, Opus 4.6, and GLM-5.3. Each run begins with the same SCoT seed harness, which we optimize on validation set \mbox{$D$} using Algorithm~\mbox{\ref{alg:metavul}}. At each iteration, we measure pairwise accuracy, precision, recall, CWE identification, and localization (Fig.~\mbox{\ref{fig:rq1-validation-metrics}}). To determine the Pareto frontier, we use pairwise accuracy and joint hit rate on the validation set. Then, without changing the harnesses, we evaluate each of the harnesses on PrimeVul: first with the model used for optimization, then with a model not used during optimization. These transfer models are Nemotron 3.5 Lightning, GLM-4.7 Flash, Opus 5, and GPT-OSS-120B (Fig.~\mbox{\ref{fig:pairwise-evolution}}).\footnote{\supp{S6_additional_results}{S6} shows each column at full width and gives a worked example of selecting the Pareto frontier.} For each evaluation, we report the best harness from the validation frontier. \supp{S5_rq1_full_metrics}{S5} reports every metric for the seed and best harness, shows how pairwise accuracy correlates with the other metrics, and shows how many harnesses beat the seed on each metric.

\textbf{Results.} Feedback-driven optimization improves validation pairwise accuracy in all four runs, with best-harness gains over the seed of $+$21.3 (Inkling-Small), $+$21.8 (DeepSeek V4 Flash), $+$23.3 (Opus 4.6) and $+$11.4 points (GLM-5.3) (Fig.~\ref{fig:rq1-validation-metrics}), and all 40 discovered harnesses beat the seed. The gains largely carry over: a frontier harness beats the seed in all 8 held-out and cross-model evaluations (Fig.~\ref{fig:pairwise-evolution}). On unseen PrimeVul data with the same model, the gains are $+$11.9, $+$25.8, $+$8.4 and $+$5.6 points in the same order; on models that took no part in optimization, they are $+$8.4 on Nemotron 3.5 Lightning, $+$7.9 on GLM-4.7 Flash, $+$14.7 on Opus 5 and $+$4.7 on GPT-OSS-120B. Pairwise accuracy generally increases with precision while recall decreases. Its relationship with joint hit rate differs across models, indicating a negative association for Inkling-Small and DeepSeek V4 Flash and a positive association for GLM-5.3 (\supptab{S5_rq1_full_metrics}{S5.2}). Only 10 of the 40 harnesses improve localization (\supptab{S5_rq1_full_metrics}{S5.3}). Consequently, these implications motivate the application of a Pareto frontier, rather than one specific harness chosen strictly on one of the criteria provided.

\begin{tcolorbox}[
    colback=blue!5,
    colframe=blue!40!black,
    boxrule=0.6pt,
    arc=4pt,
    left=4pt, right=4pt, top=2pt, bottom=2pt,
    before skip=4pt, after skip=4pt,
    breakable,
]
\textbf{RQ1 Findings.} MetaVul improves pairwise accuracy by 11.4 to 23.3 points ($1.6\times$ to $4.3\times$) over the seed, and the gains hold on unseen data (5.6 to 25.8 points, $1.3\times$ to $2.9\times$) and unseen models (4.7 to 14.7 points, $1.3\times$ to $1.7\times$). It is a critical challenge to concurrently improve discrimination, localization, and CWE identification, as singular improvements to one do not necessarily directly translate to each of the others.
\end{tcolorbox}

\input{results/related_comparison}

\section{Comparison to Other Methods}
\label{sec:related-comparison}
\label{sec:rq3}

\textbf{Motivation \& Approach.} Related work evaluates four LLM-based and agentic vulnerability detectors on PrimeVul using GPT-4o, providing a common backbone for comparison (Sec.~\ref{sec:related-work}). This motivates \textbf{RQ2:} \textbf{\emph{How does MetaVul compare with other LLM-based or agentic methods?}} Because GPT-4o was costly to evaluate through OpenRouter, our budget did not permit transferring every discovered harness. We randomly selected the DeepSeek V4 Flash discovery run and evaluated the six harnesses with the highest validation pairwise accuracy to test whether its strongest candidates transferred to GPT-4o.\footnote{At the time of evaluation, GPT-4o cost \$10 per million output tokens on OpenRouter.} We also evaluated iteration 7; iteration 13 did not complete. \supp{S7_gpt4o_transfer}{S7} reports all completed transfers, while Table~\ref{tab:primevul-llm-performance} compares the seed with iteration 12, the transferred harness with the highest accuracy.

\textbf{Results.}
MetaVul correctly discriminates 103 of 435 pairs, compared with 74 for the seed
and 81 for the next-best method, whose results are reported from prior work (Table~\ref{tab:primevul-llm-performance}). Notably, MetaVul uses a single detector agent, whereas GPTLens, VulTrial and ConfColl coordinate multiple agents. After removing 5 pairs overlapping
validation, MetaVul still correctly discriminates 100 of 430 pairs. It also
achieves the highest accuracy and the lowest false-positive rate among methods
with non-trivial recall. The GPT-4o evaluation was limited to a subset of discovered harnesses due to budget constraints. This result should not be considered an upper bound. The evaluation of additional harnesses may in fact yield stronger performance results.

\begin{tcolorbox}[
    colback=blue!5,
    colframe=blue!40!black,
    boxrule=0.6pt,
    arc=4pt,
    left=4pt, right=4pt, top=2pt, bottom=2pt,
    before skip=4pt, after skip=4pt,
]
\textbf{RQ2 Findings.} Using the same GPT-4o model, MetaVul achieves stronger pairwise accuracy than other existing vulnerability-detection methods. This includes multi-agent approaches, even after removing any validation overlap. These results are conclusive that the learned harness directly improves detection using a single detector agent, without ever changing the underlying model used.
\end{tcolorbox}

\section{Optimization Integrity}
\label{sec:rq4}

\textbf{Motivation \& Approach.} Because the optimizer receives labeled validation feedback, its gains could come from shortcuts rather than better harness design. We therefore ask \textbf{RQ3:} \textbf{\emph{Do MetaVul's gains rely on data leakage, embedded answers, or trivial verdict bias?}} We test two possible shortcut pathways. First, we verify that the optimizer cannot access PrimeVul during discovery. Second, we use GPT-OSS-120B as a judge to examine all 44 harnesses using only their text. The judge assigns numerical ratings from 0 to 4 (where 0 means missing/unusable and 4 means precise/complete) for respecting relevant constraints: (1) for preserving SCoT, (2) for avoiding prohibited content such as examples, labels, dataset hints, or verdict-bias instructions, and (3) for preserving the output format. Table~\ref{tab:shortcut-main} reports these checks for the Pareto-frontier harnesses, with full judge results in \supp{S2_harness_quality}{S2}.

\input{results/shortcut_table}

\textbf{Results.} We find no evidence that the Pareto-frontier harnesses exploit the tested shortcuts. All 11 receive $4/4$ for preserving SCoT and avoiding prohibited content. Two receive lower output-format ratings because of conflicting instructions, a quality issue rather than a shortcut. Every frontier harness reduces both false positives and the total number of flagged functions, with false positives falling by 33\% to 60\%.

\begin{tcolorbox}[
    colback=blue!5,
    colframe=blue!40!black,
    boxrule=0.6pt,
    arc=4pt,
    left=4pt, right=4pt, top=2pt, bottom=2pt,
    before skip=4pt, after skip=4pt,
    breakable,
]
\textbf{RQ3 Findings.} We find no evidence that MetaVul's Pareto-frontier harnesses rely on data leakage, embedded answers, or increased flagging. All reduce false positives while preserving the required reasoning scaffold.
\end{tcolorbox}

\section{Limitations}
\label{sec:limitations}

We identify some limitations and threats in this work. \textbf{Same model for both roles.} During harness discovery, the detection agent and the optimizer agent use the same model. Our cross-model transfer experiments help to mitigate this risk. We tested whether the discovered harnesses still improved detection rates on models that took \emph{no} part in optimization, and showed that the gains carry over, hence the gains cannot be attributed to one specific model. \textbf{Seed harness.} Although we selected a strong seed based on a pilot comparison of different reasoning strategies, ultimately, the space of possible seeds is large, so it was infeasible to test whether other seeds lead to different or better harnesses. \textbf{Contribution of workspace components.} We do not establish which parts of the optimizer's workspace (earlier harnesses, aggregate metrics, the score history, or example-by-example predictions with labels) account for the improvement. A preliminary ablation that removed these components level by level was inconclusive (\supp{S11_ablation}{S11}). \textbf{Language and granularity.} All experiments operate on C/C++ functions, so results may not carry over to memory-safe or dynamically typed languages. Detection is also performed at the function level, without any surrounding context or access to the rest of the repository. \textbf{Cost and access to infrastructure.} Access to API infrastructure was a severe limitation, although the results so far are very promising. Because both agents are launched as CLI processes (\texttt{codex exec} and \texttt{claude -p}), we mostly relied on subscriptions rather than per-token API billing to obtain these results. Subscriptions are limited by 5-hour and weekly usage windows, and a single optimization run, which evaluates every candidate harness on 404 functions, exhausts them very quickly. OpenRouter options for open-weight models were cheap (for example, \$0.03 per million input tokens and \$0.17 per million output tokens for GPT-OSS-120B; \supp{S12_inference_prices}{S12}), but GPT-4o was not, at \$2.50 per million input tokens and \$10.00 per million output tokens~\citep{openrouter_gpt4o}. Due to these previously listed constraints, we were limited by the number of feasible iterations, repeated runs, and model pairings. \textbf{Data contamination.} We cannot fully exclude that the benchmarks appeared in the models' training data. Some of the evaluated models have training cutoffs that postdate the release of the benchmarks available. However, the three following observations directly suggest that this is unlikely and mitigate this concern. GPT-4o has a training cutoff (October 2023) that precedes the release of PrimeVul, TitanVul and CWEval, and harnesses learned with DeepSeek V4 Flash still improve on it (79-101 vs.\ 71 correct pairs; \supp{S7_gpt4o_transfer}{S7}). The seed harness also scores near the floor on pairwise accuracy ($10.5\%$), which memorized fixes would not produce, and grouping all 430 PrimeVul pairs by CVE year (2008-2022) shows no monotonic trend.\footnote{We maintain that, if a model memorized fixes during training, older bugs, which have been public for longer, should be easier. Each PrimeVul pair is a real bug reported in a particular year. So, to test this, we group pairs by year (2008-2022) and measure performance in each group. We found no consistent trend across runs. Per-year results are in our \href{https://github.com/ARiSE-Lab/MetaVul/blob/main/results/cve_year/report.md}{CVE-year report}.}

\section{Future Work}
\label{sec:future-work}
There are several directions that could be targeted by future work. First, it could be worthwhile to swap or pair different optimizer and detector agents, for example a smaller detector with a stronger optimizer, or Claude Code with Codex, to measure how much each role contributes to the discovered harnesses. The trade-off between discrimination, localization, and CWE identification should also be studied directly, since techniques that improve these metrics together remain underexplored. Starting discovery from a different seed harness could also show how the seed limits or biases evolution downstream. We can also think of many enhancements to the harness itself, such as a spec-driven approach to harness design, where a declarative specification can outline the input and output format, constrain each candidate harness, and check it before evaluation. The detector could also be given other context, like pre-extracted data-flow and program dependence graphs, or call graphs and caller/callee information (using the whole project), which can be pre-computed for the data and included in the workspace. Finally, direct integration into agentic exploit generation could be explored, which has implications for improving cost and efficiency.

\section{Conclusion}
\label{sec:conclusion}
MetaVul automatically discovers vulnerability-detection harnesses that outperform a hand-designed seed. Across four runs, validation pairwise accuracy improves by 11.4-23.3 points ($1.6\times$-$4.3\times$). These gains persist across all eight held-out and cross-model evaluations (up to 25.8 points on unseen PrimeVul data and 14.7 on an unseen model). On GPT-4o, a transferred harness correctly discriminates 103 of 435 pairs, outperforming existing methods and achieving the highest accuracy. We find no evidence of shortcuts: every Pareto-frontier harness preserves SCoT, avoids prohibited content, and reduces false positives by 33\%-60\%. MetaVul thus addresses two challenges. Firstly, it \emph{reduces false alarms}, cutting false positives while improving discrimination between security-relevant changes and benign code variations. Secondly, it \emph{improves generalization}, showing that the gains hold on unseen data and unseen models rather than overfitting.

These results have further implications for real-world vulnerability detection, including, but not limited to, analyzing new codebases, automated exploit generation, and triage. First, the transferability results show that this technique can transfer to unseen codebases. Second, it can make automated exploit generation more efficient and scalable: it can filter out benign code from promising vulnerabilities before exploit generation and execution, which are computationally expensive (see Future Work, Sec.~\ref{sec:future-work}). Even after exploit generation, people must decide how serious an exploit is, what its potential downstream effects are, and how quickly a fix should be applied. Thus, there are major implications for security auditing and triage, as this technique can reduce the need for developer intervention.

\section*{Reproducibility Statement}
Our code has been publicly available at \url{https://github.com/ARiSE-Lab/MetaVul}~\citep{metavul_repo} since September 30, 2026. This work, from conception to implementation, took place between February and September 2026.

\section*{Acknowledgements}
Part of this work was supported by a Tinker Grant from Thinking Machines Lab (TML). We thank TML for their support. We also thank Ciaran Ramasastry Walsh for his discussions and early explorations in the Codex environment.

\section*{AI Use Statement}
We used AI to improve transitions and wording in our writing to make it clearer for our audience in cases where page limit constraints are a concern. We used our hand-written drafts prior to improving transitions or wording. We did not use AI to conceive any research ideas. All ideas came from reading the literature and related work, discussions among the authors, and insights from past projects, and we first wrote our own notes and human-authored drafts. We also used Claude Code and Codex for code maintenance and debugging. All AI-assisted text and code were reviewed by the authors.

\bibliography{main}
\bibliographystyle{iclr2027_conference}

\end{document}

%% file: macros.tex
\definecolor{noteAcolor}{RGB}{130, 100, 190}

\definecolor{noteAfeedbackcolor}{RGB}{160, 90, 200}

\definecolor{noteCcolor}{RGB}{51, 153, 255}

\definecolor{noteBcolor}{RGB}{0,128,128}

\newcommand{\suppurl}{https://github.com/ARiSE-Lab/MetaVul/blob/main/supplementary_materials}
\newcommand{\supp}[2]{Supp.~\href{\suppurl/#1.pdf}{#2}}
\newcommand{\supptab}[2]{Supp.~Table~\href{\suppurl/#1.pdf}{#2}}

%% file: sections/fig_pipeline.tex
\usetikzlibrary{arrows.meta,positioning,fit,backgrounds,calc,shapes.geometric}

\definecolor{searchc}{HTML}{2F6DB5}  
\definecolor{transc}{HTML}{D9822B}   
\definecolor{optc}{HTML}{7B4FA0}     
\definecolor{okc}{HTML}{3A9A5B}      
\definecolor{badc}{HTML}{C8453B}     
\providecolor{optdeep}{HTML}{4A2472}  
\providecolor{detdeep}{HTML}{1A3F7A}  

\tikzset{
  harnessicon/.pic={
    \foreach \o in {0.12,0.06,0} {
      \draw[black!60, fill=white] ($(-0.25,-0.3)+(\o,\o)$) rectangle ++(0.4,0.5);
    }
    \foreach \y in {-0.18,-0.08,0.02,0.12} {
      \draw[black!35] (-0.19,\y) -- (0.09,\y);
    }
  },
  metricsicon/.pic={
    \draw[black!60] (-0.3,0.25) -- (-0.3,-0.25) -- (0.3,-0.25);
    \fill[searchc!55] (-0.22,-0.25) rectangle (-0.08,-0.05);
    \fill[searchc!75] (-0.03,-0.25) rectangle (0.11,0.18);
    \fill[searchc!40] (0.16,-0.25) rectangle (0.28,0.05);
  },
  historyicon/.pic={
    \draw[black!60, fill=white, line width=0.6pt] (0,0) circle (0.25);
    \draw[black!70, line width=0.7pt] (0,0) -- (0,0.16);
    \draw[black!70, line width=0.7pt] (0,0) -- (0.12,0);
    \fill[black!70] (0,0) circle (0.025);
  },
  predicticon/.pic={
    \foreach \y/\c in {0.14/okc,0/badc,-0.14/okc} {
      \fill[black!15] (-0.3,\y-0.045) rectangle (0.12,\y+0.045);
      \fill[\c] (0.22,\y) circle (0.055);
    }
  },
  lockicon/.pic={
    \draw[black!65, line width=0.8pt] (-0.07,0) -- (-0.07,0.06) arc (180:0:0.07) -- (0.07,0);
    \fill[black!65, rounded corners=0.4pt] (-0.11,-0.13) rectangle (0.11,0.02);
  },
}
\newcommand{\agentlogo}[1]{\raisebox{-0.3\height}{\includegraphics[height=1.7em]{figures/logos/#1}}}
\newcommand{\stepnum}[1]{\tikz[baseline=(sn.base)]\node[circle, draw=black!75, fill=white, line width=0.6pt,
  inner sep=0.5pt, minimum size=1.55em, font=\sffamily\bfseries\scriptsize] (sn) {#1};}

\begin{figure}[t]
\centering
\resizebox{\textwidth}{!}{%
\begin{tikzpicture}[
  font=\sffamily\small,
  >={Stealth[length=2.2mm]},
  phase/.style={draw, rounded corners=3pt, line width=0.8pt,
                minimum width=6.9cm, minimum height=2.5cm},
  ptitle/.style={anchor=north west, align=left, inner sep=0pt},
  pbody/.style={anchor=south west, align=left, inner sep=0pt, font=\sffamily\footnotesize},
  data/.style={cylinder, shape border rotate=90, aspect=0.22, draw=black!55,
               cylinder uses custom fill, cylinder body fill=#1!8,
               cylinder end fill=#1!22, minimum width=2.1cm, minimum height=1.95cm,
               align=center, inner sep=1pt, font=\sffamily\footnotesize},
  stage/.style={draw=searchc, fill=searchc!6, rounded corners=2pt, align=center,
                inner sep=4pt, text width=3.6cm, minimum height=2.6cm,
                font=\sffamily\footnotesize},
  hcard/.style={draw=black!70, fill=white, minimum width=0.6cm,
                minimum height=0.75cm, inner sep=0pt, font=\sffamily\scriptsize},
  lbl/.style={font=\sffamily\scriptsize, text=black!70, align=center},
  icap/.style={font=\sffamily\scriptsize, text=black!75, inner sep=1pt},
  flow/.style={->, line width=0.8pt},
]

\node[font=\sffamily\bfseries, anchor=west] at (-6.0,3.2) {System Overview};

\node[phase, draw=searchc, fill=searchc!5] (val) at (0,0) {};
\node[ptitle] at ($(val.north west)+(0.2,-0.2)$)
  {\stepnum{1}\; \textbf{Validation}\\ {\footnotesize\itshape harness discovery}};
\node[pbody] at ($(val.south west)+(0.2,0.25)$)
  {Optimizer: \textcolor{optdeep}{$\mathcal{A}_{\mathrm{opt}}$} $+\,M_s$\\ Detector: \textcolor{detdeep}{$\mathcal{A}_{\mathrm{det}}$} $+\,M_s$};
\node[data=searchc, anchor=east] at ($(val.east)+(-0.2,0)$)
  {\textbf{TitanVul}\\ \textbf{+ CWEval}\\ 202 pairs ($D$)};

\node[lbl, left=0.7cm of val] (seed) {seed $h_0$};
\draw[flow] (seed) -- (val);

\node[hcard] (hc2) at (5.2,0.20) {};
\node[hcard] (hc1) at (5.05,0.10) {};
\node[hcard] (hc0) at (4.9,0.00) {$h_t$};
\node[lbl, below=0.3cm of hc1] (hlab) {$\mathcal{H}=\{h_0,\dots,h_T\}$\\frozen};
\coordinate (Heast) at ($(hc2.east)+(0.1,-0.1)$);

\draw[flow, searchc] (val.east) -- (hc0.west)
  node[midway, above, lbl] {outputs};

\node[phase, draw=searchc, fill=searchc!5] (ho) at (11.3,1.5) {};
\node[ptitle] at ($(ho.north west)+(0.2,-0.2)$)
  {\stepnum{2}\; \textbf{Held-out evaluation}\\ {\footnotesize\itshape unseen dataset transfer}};
\node[pbody] at ($(ho.south west)+(0.2,0.25)$)
  {\textcolor{detdeep}{$\mathcal{A}_{\mathrm{det}}$} $+\,M_s$};
\node[data=black, anchor=east] at ($(ho.east)+(-0.2,0)$)
  {\textbf{PrimeVul}\\ 430 pairs\\ unseen};

\node[phase, draw=transc, fill=transc!6] (cm) at (11.3,-1.5) {};
\node[ptitle] at ($(cm.north west)+(0.2,-0.2)$)
  {\stepnum{3}\; \textbf{Cross-model transfer}\\ {\footnotesize\itshape unseen model transfer}};
\node[pbody] at ($(cm.south west)+(0.2,0.25)$)
  {\textcolor{detdeep}{$\mathcal{A}_{\mathrm{det}}$} $+\,M_t$};
\node[data=black, anchor=east] at ($(cm.east)+(-0.2,0)$)
  {\textbf{PrimeVul}\\ 430 pairs\\ unseen};

\draw[flow] (Heast) -- (ho.west);
\draw[flow] (Heast) -- (cm.west);

\draw[black!45, dashed, line width=0.7pt] (7.25,2.85) -- (7.25,-2.85)
  node[pos=0.0, above, lbl] {no feedback};
\node[lbl, fill=white, inner sep=1.5pt] at (7.25,0) {replay\\all $h_t$};

\node[stage] (ws) at (-3.4,-6.1) {};
\node[anchor=north, inner sep=0pt, font=\sffamily\footnotesize\bfseries]
  at ($(ws.north)+(0,-0.15)$) {Workspace $W_t$};
\pic[scale=1.25] at ($(ws.center)+(-0.85,0.35)$) {harnessicon};
\node[icap] at ($(ws.center)+(-0.85,-0.15)$) {harnesses};
\pic[scale=1.25] at ($(ws.center)+(0.85,0.35)$) {metricsicon};
\node[icap] at ($(ws.center)+(0.85,-0.15)$) {metrics};
\pic[scale=1.25] at ($(ws.center)+(-0.85,-0.62)$) {historyicon};
\node[icap] at ($(ws.center)+(-0.85,-1.07)$) {history};
\pic[scale=1.25] at ($(ws.center)+(0.85,-0.62)$) {predicticon};
\node[icap] at ($(ws.center)+(0.85,-1.07)$) {preds + labels};

\node[stage, draw=optc, fill=optc!7] (opt) at (1.3,-6.1)
  {\textbf{Optimizer agent}\\ \textcolor{optdeep}{$\mathcal{A}_{\mathrm{opt}}$} $+\,M_s$\\[3pt]
   {\scriptsize\agentlogo{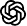}\,Codex\quad\agentlogo{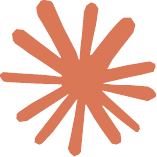}\,Claude Code}\\[3pt]
   \textsc{Propose}: rewrites\\ $h_{t-1} \to h_t$};
\node[stage] (ev) at (6.0,-6.1)
  {\textbf{Detection agent}\\ \textcolor{detdeep}{$\mathcal{A}_{\mathrm{det}}$} $+\,M_s$\\[3pt]
   {\scriptsize\agentlogo{openai}\,Codex\quad\agentlogo{claude}\,Claude Code}\\[3pt]
   \textsc{Evaluate}: runs $h_t$ on $D$\\ $\to (m_t, z_t)$};
\node[stage] (sel) at (10.7,-6.1)
  {\textbf{Select}\\[2pt] $s_t = \textsc{PairwiseAcc}(z_t)$\\ if $s_t > s_{\mathrm{best}}$: $h_{\mathrm{best}} \gets h_t$\\ else: $c \gets c+1$};

\draw[flow] (ws) -- (opt);
\draw[flow] (opt) -- (ev);
\draw[flow] (ev) -- (sel);

\coordinate (looptop) at ($(ws.north)+(0,0.7)$);
\draw[flow] (sel.north) |- (looptop) -- (ws.north);

\coordinate (ret) at ($(looptop)!0.5!(sel.north |- looptop)$);
\fill[white, rounded corners=2pt] ($(ret)+(-2.05,-0.3)$) rectangle ($(ret)+(2.05,0.3)$);
\foreach \dx/\ic in {-1.5/harnessicon, -0.5/metricsicon, 0.5/predicticon, 1.5/historyicon} {
  \pic[scale=0.8] at ($(ret)+(\dx,0)$) {\ic};
}
\node[lbl, above=0.32cm of ret] (retlab)
  {$\mathcal{H} \gets \mathcal{H} \cup \{(h_t, m_t, z_t)\}$: harness, metrics, predictions, history back to $W_{t+1}$};

\node[stage, draw=black!60, fill=black!3, minimum height=1.5cm, text width=4.4cm, below=1.1cm of sel] (par)
  {\textbf{Pareto filter}\\ $\mathcal{F} \gets \textsc{ParetoFront}(\mathcal{H})$\\ on pairwise acc.\\ and joint hit rate};
\draw[flow] (sel.south) -- (par.north)
  node[midway, right, lbl, align=left] {stop: $t = T$ (15)\\ or $c \ge K$ (4)};

\node[lbl, left=0.9cm of par, font=\sffamily\footnotesize] (out)
  {return $\mathcal{F}$:\\ Pareto-optimal harnesses};
\draw[flow] (par) -- (out);

\foreach \s/\n in {ws/1a, opt/1b, ev/1c, sel/1d, par/1e} {
  \node[inner sep=0pt] at ($(\s.north west)+(0.32,-0.32)$) {\stepnum{\n}};
}

\begin{scope}[on background layer]
  \node[draw=searchc!70, dashed, rounded corners=4pt, line width=0.8pt,
        fit=(ws)(opt)(ev)(sel)(par)(out)(looptop)(retlab), inner sep=8pt,
        label={[font=\sffamily\bfseries, anchor=north west]south west:Inside \stepnum{1}: harness discovery loop}]
        (zoom) {};
  \foreach \s/\sbcap in {opt/{reads only its workspace},
                       ev/{empty workspace}} {
    \node[draw=black!55, fill=black!6, rounded corners=4pt, line width=0.9pt,
          fit=(\s), inner sep=4pt] (sb-\s) {};
    \pic at ($(sb-\s.south west)+(0.3,-0.3)$) {lockicon};
    \node[icap, anchor=west, align=left] at ($(sb-\s.south west)+(0.45,-0.33)$)
      {\textbf{sandbox}: \sbcap};
  }
\end{scope}

\draw[searchc!60, dashed] (val.south west) -- (zoom.north west);
\draw[searchc!60, dashed] (val.south east) -- ($(zoom.north west)!0.55!(zoom.north east)$);

\end{tikzpicture}%
}
\caption{\textbf{MetaVul pipeline.}
(1) Starting from seed $h_0$, harnesses are discovered on the validation set.
In each iteration, (1a) the workspace stores previous harnesses and results, (1b) $\mathcal{A}_{\mathrm{opt}}$ proposes a new harness $h_t$, (1c) $\mathcal{A}_{\mathrm{det}}$ evaluates it, and (1d) the best result is selected.
The Pareto filter (1e) retains harnesses based on pairwise accuracy and joint hit rate.
Each harness is then frozen and tested on unseen PrimeVul using (2) the search model $M_s$ and (3) a different transfer model $M_t$.
PrimeVul provides no optimization feedback.}
\label{fig:pipeline}
\end{figure}

%% file: results/pairwise_evolution_data.tex
\definecolor{evoAVal}{RGB}{25,75,145}      
\definecolor{evoAHeld}{RGB}{0,135,170}     
\definecolor{evoAXfer}{RGB}{105,55,155}    
\definecolor{evoBVal}{RGB}{44,160,44}       
\definecolor{evoBHeld}{RGB}{0,110,100}     
\definecolor{evoBXfer}{RGB}{140,86,75}      
\definecolor{evoCVal}{RGB}{255,127,14}      
\definecolor{evoCHeld}{RGB}{214,39,40}      
\definecolor{evoCXfer}{RGB}{227,119,194}    
\definecolor{evoDVal}{RGB}{127,127,127}     
\definecolor{evoDHeld}{RGB}{20,40,110}      
\definecolor{evoDXfer}{RGB}{166,124,0}      

\pgfplotsset{
    evo/series/.style={line width=0.65pt, mark size=1.3pt,
                       mark options={solid, draw=white, line width=0.25pt, fill=#1}},
    evo/aval/.style={evoAVal, evo/series=evoAVal, solid, mark=*},
    evo/aheld/.style={evoAHeld, evo/series=evoAHeld, solid, mark=square*},
    evo/axfer/.style={evoAXfer, evo/series=evoAXfer, solid, mark=triangle*, mark size=1.6pt},
    evo/bval/.style={evoBVal, evo/series=evoBVal, solid, mark=*},
    evo/bheld/.style={evoBHeld, evo/series=evoBHeld, solid, mark=square*},
    evo/bxfer/.style={evoBXfer, evo/series=evoBXfer, solid, mark=triangle*, mark size=1.6pt},
    evo/cval/.style={evoCVal, evo/series=evoCVal, solid, mark=*},
    evo/cheld/.style={evoCHeld, evo/series=evoCHeld, solid, mark=square*},
    evo/cxfer/.style={evoCXfer, evo/series=evoCXfer, solid, mark=triangle*, mark size=1.6pt},
    evo/dval/.style={evoDVal, evo/series=evoDVal, solid, mark=*},
    evo/dheld/.style={evoDHeld, evo/series=evoDHeld, solid, mark=square*},
    evo/dxfer/.style={evoDXfer, evo/series=evoDXfer, solid, mark=triangle*, mark size=1.6pt},
    evo/bestpair/.style={only marks, mark=star, mark size=2.8pt, line width=0.7pt, mark options={solid}},
    evo/block/.style={
        scale only axis, height=2.2cm,  
        tick label style={font=\footnotesize},
        label style={font=\small},
        title style={font=\footnotesize, align=center},
        ymajorgrids, grid style={gray!20},
        unbounded coords=jump,
        clip mode=individual,  
    },
    evo/pair/.style={ymin=0, ymax=42},
    evo/loc/.style={ymin=10, ymax=45},
    evo/hit/.style={ymin=10, ymax=80, ytick={20,40,60,80}},
    evo/joint/.style={ymin=0, ymax=36, ytick={0,10,20,30}},
    evo/legend/.style={legend style={at={(0.5,-0.45)}, anchor=north, legend columns=1,
                       font=\fontsize{7.5}{8.5}\selectfont, draw=none, fill=none, cells={align=left}, row sep=-1pt},
                       legend cell align=left,
                       legend image post style={line width=1.6pt, every mark/.append style={scale=1.6}},
                       legend image code/.code={\draw[mark repeat=2, mark phase=2] plot coordinates {(0cm,0cm) (0.2cm,0cm) (0.4cm,0cm)};}},
    evo/claudex/.style={width=2.1cm, xmin=-0.45, xmax=4.5, xtick={0,...,4}},
    evo/glmx/.style={width=2.9cm, xmin=-0.6, xmax=8.5, xtick={0,2,...,8}, minor xtick={0,...,8}},
    evo/codexx/.style={width=3.45cm, xmin=-2.4, xmax=15.5, xtick={0,5,10,15}, minor xtick={0,...,15}},
    evo/pending/.style={xmin=-1.2, xmax=4.5, xtick={0,...,4}},
}
\tikzset{evolbl/.style={font=\fontfamily{phv}\selectfont\tiny\bfseries, inner xsep=0.4pt, inner ysep=0.2pt,
                        outer sep=0pt, text depth=0pt, rounded corners=0.6pt, line width=0.2pt}}
\newcommand{\evonum}[1]{\scalebox{0.85}{#1}}
\DeclareRobustCommand{\capband}{\tikz[baseline=-0.55ex]\fill[yellow!30] (0,-0.13) rectangle (0.42,0.13);}
\DeclareRobustCommand{\capstar}{\tikz[baseline=-0.6ex]\node[star, star points=5, star point ratio=2.25, draw=black!70, fill=black!15, inner sep=0.9pt, line width=0.5pt]{};}
\newcommand{\evoband}[1]{\fill[yellow!30] ({axis cs:#1-0.35,0}|-{rel axis cs:0,0}) rectangle ({axis cs:#1+0.35,0}|-{rel axis cs:0,1});}
\pgfplotsset{evo/noprompt/.style={black, densely dotted, line width=0.5pt, mark=none}}
\newcommand{\evopending}{\node[font=\footnotesize\itshape, text=gray] at (rel axis cs:0.5,0.5) {to be added};}

\pgfplotstableread{
iter valpair heldpair xferpair valloc heldloc xferloc valhit heldhit xferhit valjoint heldjoint xferjoint
0    10.40   10.47    22.56    36.4   24.0    24.5     68.5    45.8    60.6     24.4    14.1    19.9
1    29.21   15.12    33.26    40.8   29.0    31.8     70.5    43.5    54.3     30.9    17.7    20.1
2    31.19   18.84    32.09    39.9   28.4    27.7     65.7    42.0    45.8     31.2    17.6    17.3
3    33.66   17.67    37.21    39.8   27.8    30.3     67.3    41.7    48.3     28.6    18.2    18.2
4    32.18   21.86    35.12    39.4   29.0    29.4     64.1    42.4    50.3     24.2    16.8    18.9
}\evoClaudeA
\pgfplotstableread{
iter valpair heldpair xferpair valloc heldloc xferloc valhit heldhit xferhit valjoint heldjoint xferjoint
0     19.80   18.37    14.42     32.3  23.0    21.4      73.0    62.6    46.5     21.9    19.8    7.0
1     25.74   25.58    15.35     35.5  26.7    23.2      61.3    42.2    40.9     21.3    16.3    11.5
2     27.23   22.33    16.98     33.5  25.1    18.3      66.2    55.8    33.8     25.2    13.4    10.9
3     28.71   24.19    15.58     35.7  27.5    17.6      65.6    54.8    34.5     25.3    17.4    9.0
4     31.19   21.16    15.35     32.6  23.8    18.1      72.5    62.9    38.4     29.8    20.9    7.9
5     26.24   22.56    14.42     32.3  25.4    18.7      66.4    51.5    36.5     27.8    17.4    9.8
6     31.19   23.95    19.07     37.4  24.7    16.9      71.5    55.8    31.9     34.1    18.9    7.9
7     27.72   22.33    14.42     34.9  26.6    17.0      63.4    48.5    27.6     27.2    18.4    8.5
8     28.71   22.79    20.70     30.7  27.3    15.5      58.9    48.5    26.2     29.2    17.8    5.6
}\evoClaudeB
\pgfplotstableread{
iter valpair heldpair xferpair valloc heldloc xferloc valhit heldhit xferhit valjoint heldjoint xferjoint
0    6.44    8.84     12.79    33.0   22.2    19.8     61.7    48.1    43.5     15.0    12.1     7.2
1    11.39   8.84     9.07     32.6   20.0    15.8     62.5    44.0    27.6     13.4    10.4     5.7
2    15.84   15.35    17.21    31.9   20.3    16.3     57.6    42.3    27.2     13.0    11.9     7.7
3    12.87   10.47    13.26    28.9   19.4    18.7     53.9    44.9    31.7     12.5    12.1     5.9
4    17.33   18.60    20.23    32.2   22.3    14.9     67.8    52.6    26.0     16.7    12.6     3.8
5    14.85   19.07    20.70    28.2   19.1    14.9     58.2    47.0    26.4     14.4    13.2     5.5
6    20.79   15.58    19.07    31.3   21.2    17.4     66.9    49.1    32.9     19.3    12.6    10.2
7    15.84   19.30    19.77    29.4   22.8    16.8     62.8    58.9    30.9     14.1    16.5     7.6
8    23.27   18.60    14.88    33.9   22.5    17.9     74.8    58.5    34.8     12.2    14.9     8.1
9    20.79   23.95    16.98    33.7   22.3    17.5     74.6    57.1    32.3      6.8    14.3     8.3
10   20.79   19.07    19.30    30.9   22.2    15.9     67.7    55.9    29.7      8.8    14.0     7.8
11   24.26   20.00    19.53    28.6   19.9    17.3     66.1    53.4    33.2      9.1    11.9     9.7
12   27.72   20.70    21.16    28.5   19.9    16.6     59.7    47.0    29.2      9.4    10.4     8.8
13   23.27   23.02    18.14    29.7   21.5    17.6     60.0    50.0    32.4      8.1    11.8     8.4
14   24.75   20.93    20.70    30.6   18.4    14.2     61.2    42.8    25.6      6.2     9.2     5.7
15   21.78   20.93    16.98    27.7   19.9    14.8     62.2    47.2    30.4      6.8     9.3     7.2
}\evoCodexA
\pgfplotstableread{
iter valpair heldpair xferpair valloc heldloc xferloc valhit heldhit xferhit valjoint heldjoint xferjoint
0      14.36   13.26   15.81    32.0    22.1    23.9    69.1    60.2    57.3     22.8    15.8    16.5
1      24.26   24.88   23.72    35.7    28.7    14.6    63.9    55.9    31.3     28.2    19.4    3.0
2        nan     nan     nan     nan     nan     nan     nan     nan     nan     nan    nan    nan
3        nan     nan     nan     nan     nan     nan     nan     nan     nan     nan    nan    nan
4      22.28   21.86   27.91    27.6    21.1    18.6    50.4    32.0    31.7     25.8    11.9    7.8
5      28.71   19.77   19.53    29.1    19.0    17.1    52.6    31.2    31.2     24.2    10.4    4.9
6      22.28   21.63   24.65    24.4    19.5    15.8    46.1    30.9    28.1     25.0    10.3    6.7
7      26.73   24.88   24.88    25.2    19.2    16.0    45.3    30.3    27.9     17.9    9.9    7.0
8      33.66   25.58   18.84    25.2    19.8    10.3    43.5    28.9    17.6     15.3    13.4    3.7
9      30.20   39.07   16.74    25.9    21.6    12.7    49.5    32.7    20.8     22.1    12.8    3.3
10     32.18   26.51   17.67    29.0    20.4    11.9    48.2    30.4    21.8     17.3    11.4    5.2
11     36.14   28.60   14.65    28.6    19.7    13.3    51.6    30.6    24.5     20.9    14.0    5.4
12     34.65   25.35   19.07    25.8    17.0    11.9    45.5    25.9    22.5     20.2    9.9    3.5
13     31.68   26.05   18.37    28.0    19.6    13.9    50.0    30.8    24.3     20.8    10.6    5.9
14     33.66   23.02   16.05    24.9    18.2    12.2    45.4    28.9    21.5     21.5    11.8    3.2
15     23.76   26.51   18.60    29.8    19.7    14.3    50.0    30.1    22.1     19.4    11.9    5.7
}\evoCodexB

%% file: results/pareto_scatter.tex
\pgfplotstableread{
iter pair joint cwe acc prec rec fone
0      6.44  15.00  21.70  51.98  51.06  95.54  66.55
1     11.39  13.39  17.90  52.97  51.69  91.09  65.95
2     15.84  12.96  20.40  54.21  52.52  87.62  65.68
3     12.87  12.50  15.20  54.46  52.60  90.10  66.42
4     17.33  16.67  21.10  55.33  53.71  75.62  62.81
5     14.85  14.43  17.50  53.22  52.05  81.68  63.58
6     20.79  19.28  24.10  54.36  53.05  69.85  60.30
7     15.84  14.12  20.00  52.24  51.89  67.82  58.80
8     23.27  12.22  16.70  55.45  54.30  68.81  60.70
9     20.79   6.76   9.50  54.84  54.63  58.42  56.46
10    20.79   8.75  13.80  53.35  52.77  61.69  56.88
11    24.26   9.09  14.30  56.47  55.86  61.69  58.63
12    27.72   9.41  15.30  58.85  57.20  69.50  62.75
13    23.27   8.11  12.20  57.36  56.93  57.79  57.36
14    24.75   6.17  11.10  56.11  54.89  64.82  59.45
15    21.78   6.76  10.80  55.72  55.35  59.20  57.21
}\valA
\pgfplotstableread{
iter pair joint cwe acc prec rec fone
0     14.36  22.81  29.80  55.20  53.08  89.60  66.67
1     24.26  28.21  41.00  59.31  60.67  53.47  56.84
2       nan    nan    nan    nan    nan    nan    nan
3       nan    nan    nan    nan    nan    nan    nan
4     22.28  25.77  34.00  59.55  57.89  70.79  63.70
5     28.71  24.18  29.70  61.19  60.27  66.83  63.38
6     22.28  25.00  35.70  55.94  55.83  56.93  56.37
7     26.73  17.92  23.60  62.13  58.42  84.16  68.97
8     33.66  15.28  26.40  61.63  63.91  53.47  58.22
9     30.20  22.08  36.40  61.63  63.43  54.95  58.89
10    32.18  17.33  30.70  62.13  63.58  58.20  60.77
11    36.14  20.93  26.70  65.10  65.97  62.38  64.12
12    34.65  20.24  31.00  64.11  65.08  60.89  62.92
13    31.68  20.78  28.60  62.94  66.25  52.74  58.73
14    33.66  21.52  32.90  61.63  62.30  58.91  60.56
15    23.76  19.35  27.40  58.71  59.63  53.93  56.64
}\valB
\pgfplotstableread{
iter pair joint cwe acc prec rec fone
0     10.40  24.37  32.80  54.21  52.38  92.57  66.91
1     29.21  30.86  37.00  63.37  65.88  55.45  60.22
2     31.19  31.25  40.60  63.61  62.22  69.31  65.57
3     33.66  28.57  36.70  65.26  63.36  72.77  67.74
4     32.18  24.24  31.30  64.36  62.50  71.78  66.82
}\valC
\pgfplotstableread{
iter pair joint cwe acc prec rec fone
0     19.80  21.90  27.60  57.18  54.88  80.69  65.33
1     25.74  21.28  28.70  59.45  58.30  67.82  62.70
2     27.23  25.23  32.70  61.44  58.58  78.11  66.95
3     28.71  25.29  33.30  62.03  62.24  60.70  61.46
4     31.19  29.79  35.10  61.63  60.93  64.85  62.83
5     26.24  27.78  34.40  62.11  60.95  66.32  63.52
6     31.19  34.09  40.90  63.50  64.06  61.50  62.76
7     27.72  27.17  35.90  60.10  59.55  66.84  62.98
8     28.71  29.21  38.20  60.79  60.78  61.39  61.08
}\valD
\definecolor{vmPair}{RGB}{30,90,180}     
\definecolor{vmJoint}{RGB}{214,110,20}   
\definecolor{vmCwe}{RGB}{190,40,60}      
\definecolor{vmAcc}{RGB}{0,130,135}      
\definecolor{vmPrec}{RGB}{44,140,60}     
\definecolor{vmRec}{RGB}{125,80,165}     
\definecolor{vmFone}{RGB}{95,95,95}      
\pgfplotsset{
    vm/line/.style={line width=0.65pt, solid, mark size=1.1pt,
                    mark options={solid, draw=white, line width=0.2pt, fill=#1}},
    vm/pair/.style={vmPair, vm/line=vmPair, mark=*},
    vm/joint/.style={vmJoint, vm/line=vmJoint, mark=square*},
    vm/cwe/.style={vmCwe, vm/line=vmCwe, mark=triangle*, mark size=1.35pt},
    vm/acc/.style={vmAcc, vm/line=vmAcc, mark=*},
    vm/prec/.style={vmPrec, vm/line=vmPrec, mark=triangle*, mark size=1.35pt},
    vm/rec/.style={vmRec, vm/line=vmRec, mark=diamond*, mark size=1.4pt},
    vm/fone/.style={vmFone, vm/line=vmFone, mark=square*},
    vm/axis/.style={ymin=0, ymax=100, ytick={0,20,...,100}},
}
\begin{figure}[!t]
\centering
\begin{tikzpicture}
\begin{groupplot}[group style={group size=4 by 1, horizontal sep=0.3cm,
                               yticklabels at=edge left, ylabels at=edge left},
    scale only axis, height=2.5cm,
    xlabel={Iteration}, ylabel={Score (\%)},
    tick label style={font=\footnotesize}, label style={font=\small},
    title style={font=\footnotesize, align=center},
    ymajorgrids, grid style={gray!20}, unbounded coords=jump, clip mode=individual]
\nextgroupplot[vm/axis, evo/codexx, title={(a) Codex + Inkling-Small}, title style={xshift=-0.2cm}, legend to name=vmlegend, legend columns=4, legend style={font=\footnotesize, draw=none, /tikz/every even column/.append style={column sep=10pt}}, legend image post style={line width=1.4pt}]
\evoband{6}\evoband{8}\evoband{12}
\addplot[vm/pair] table[x=iter, y=pair] {\valA}; \addlegendentry{Pairwise accuracy}
\addplot[vm/joint] table[x=iter, y=joint] {\valA}; \addlegendentry{Joint hit rate}
\addplot[vm/prec] table[x=iter, y=prec] {\valA}; \addlegendentry{Precision}
\addplot[vm/rec] table[x=iter, y=rec] {\valA}; \addlegendentry{Recall}
\addplot[evo/bestpair, color=vmPair] coordinates {(12,27.72)};
\nextgroupplot[vm/axis, evo/codexx, title={(b) Codex + DS-V4-Flash}]
\evoband{1}\evoband{5}\evoband{9}\evoband{11}\evoband{14}
\addplot[evo/noprompt, forget plot] coordinates {(1,24.26) (4,22.28)};
\addplot[evo/noprompt, forget plot] coordinates {(1,28.21) (4,25.77)};
\addplot[evo/noprompt, forget plot] coordinates {(1,60.67) (4,57.89)};
\addplot[evo/noprompt, forget plot] coordinates {(1,53.47) (4,70.79)};
\addplot[vm/pair] table[x=iter, y=pair] {\valB};
\addplot[vm/joint] table[x=iter, y=joint] {\valB};
\addplot[vm/prec] table[x=iter, y=prec] {\valB};
\addplot[vm/rec] table[x=iter, y=rec] {\valB};
\addplot[evo/bestpair, color=vmPair] coordinates {(11,36.14)};
\nextgroupplot[vm/axis, evo/claudex, title={(c) CC + Opus 4.6}, title style={xshift=0.05cm}]
\evoband{2}\evoband{3}
\addplot[vm/pair] table[x=iter, y=pair] {\valC};
\addplot[vm/joint] table[x=iter, y=joint] {\valC};
\addplot[vm/prec] table[x=iter, y=prec] {\valC};
\addplot[vm/rec] table[x=iter, y=rec] {\valC};
\addplot[evo/bestpair, color=vmPair] coordinates {(3,33.66)};
\nextgroupplot[vm/axis, evo/glmx, title={(d) CC + GLM-5.3}]
\evoband{6}
\addplot[vm/pair] table[x=iter, y=pair] {\valD};
\addplot[vm/joint] table[x=iter, y=joint] {\valD};
\addplot[vm/prec] table[x=iter, y=prec] {\valD};
\addplot[vm/rec] table[x=iter, y=rec] {\valD};
\addplot[evo/bestpair, color=vmPair] coordinates {(4,31.19) (6,31.19)};
\end{groupplot}
\end{tikzpicture}

\vspace{1pt}
\ref*{vmlegend}\par\vspace{2pt}
{\footnotesize\setlength{\fboxsep}{2.5pt}\fcolorbox{gray!45}{white}{%
\tikz[baseline=-0.55ex]\fill[yellow!30] (0,-0.13) rectangle (0.42,0.13);\ Pareto front\qquad
\tikz[baseline=-0.55ex]\draw[black, densely dotted, line width=0.5pt] (0,0) -- (0.42,0);\ No metric\qquad
\tikz[baseline=-0.55ex]\node[star, star points=5, star point ratio=2.25, draw=vmPair, fill=vmPair!15, inner sep=0.9pt, line width=0.5pt]{};\ Best pairwise accuracy}}
\caption[Validation metrics across optimization]{\textbf{Validation metrics across optimization.} Each panel shows pairwise accuracy, joint hit rate, precision, and recall on TitanVul+CWEval for one agent-model configuration (CC = Claude Code; DS = DeepSeek).\protect\footnotemark{} Pairwise accuracy generally rises with precision while recall declines, indicating fewer false alarms but more missed vulnerabilities. Joint hit rate varies by model. \capband{} marks the Pareto front and \capstar{} the best pairwise accuracy.}
\label{fig:rq1-validation-metrics}
\end{figure}
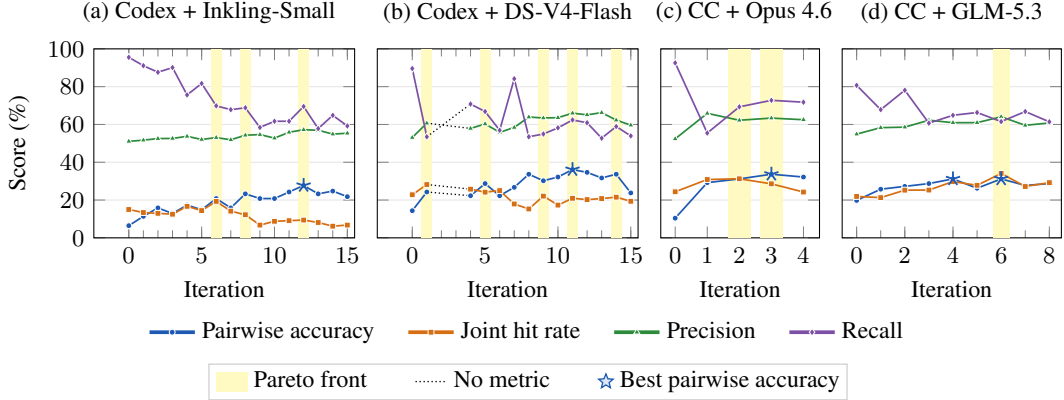
\footnotetext{We exclude accuracy, F1, and CWE accuracy as accuracy closely follows precision, F1 incorporates findings from precision and recall, while CWE accuracy overlaps with Joint. \supptab{S5_rq1_full_metrics}{S5.1} reports all metrics.}

%% file: results/pairwise_evolution_plot.tex

\begin{figure}[!tb]
\centering
\begin{tikzpicture}
\begin{groupplot}[group style={group size=4 by 2, horizontal sep=0.3cm, vertical sep=0.5cm,
                               xlabels at=edge bottom, xticklabels at=edge bottom,
                               ylabels at=edge left, yticklabels at=edge left},
                  evo/block, xlabel={Iteration}]
\nextgroupplot[evo/pair, evo/codexx, title={(a) Codex + Inkling-Small}, title style={xshift=-0.2cm}, ylabel={Pairwise acc.\ (\%)}]
\evoband{6}\evoband{8}\evoband{12}
\addplot[evo/aval]  table[x=iter, y=valpair]  {\evoCodexA};
\addplot[evo/aheld] table[x=iter, y=heldpair] {\evoCodexA};
\addplot[evo/axfer] table[x=iter, y=xferpair] {\evoCodexA};
\addplot[evo/bestpair, color=evoAVal]  coordinates {(12,27.72)};
\addplot[evo/bestpair, color=evoAHeld] coordinates {(9,23.95)};
\addplot[evo/bestpair, color=evoAXfer] coordinates {(12,21.16)};
\node[evolbl, text=evoAXfer, fill=white, draw=evoAXfer!40, anchor=south, yshift=2.5pt] at (axis cs:0,12.79) {\evonum{12.8}};
\node[evolbl, text=evoAHeld, fill=white, draw=evoAHeld!40, anchor=east, xshift=-2.5pt] at (axis cs:0,8.84) {\evonum{8.8}};
\node[evolbl, text=evoAVal,  fill=white, draw=evoAVal!40,  anchor=north, yshift=-2.5pt] at (axis cs:0,6.44) {\evonum{6.4}};
\node[evolbl, text=evoAVal,  fill=evoAVal!3,  draw=evoAVal!40,  anchor=south, yshift=3pt] at (axis cs:12,27.72) {\evonum{27.7}};
\node[evolbl, text=evoAHeld, fill=evoAHeld!3, draw=evoAHeld!40, anchor=south, yshift=3pt] at (axis cs:9,23.95)  {\evonum{24.0}};
\node[evolbl, text=evoAXfer, fill=evoAXfer!3, draw=evoAXfer!40, anchor=south east, xshift=-1.5pt, yshift=1.5pt] at (axis cs:12,21.16) {\evonum{21.2}};
\nextgroupplot[evo/pair, evo/codexx, title={(b) Codex + DS-V4-Flash}]
\evoband{1}\evoband{5}\evoband{9}\evoband{11}\evoband{14}
\addplot[evo/noprompt] coordinates {(1,24.26) (4,22.28)};
\addplot[evo/noprompt, forget plot] coordinates {(1,23.72) (4,27.91)};
\addplot[evo/noprompt, forget plot] coordinates {(1,24.88) (4,21.86)};
\addplot[evo/bval]  table[x=iter, y=valpair]  {\evoCodexB};
\addplot[evo/bheld] table[x=iter, y=heldpair] {\evoCodexB};
\addplot[evo/bxfer] table[x=iter, y=xferpair] {\evoCodexB};
\addplot[evo/bestpair, color=evoBVal]  coordinates {(11,36.14)};
\addplot[evo/bestpair, color=evoBHeld] coordinates {(9,39.07)};
\addplot[evo/bestpair, color=evoBXfer] coordinates {(4,27.91)};

\node[evolbl, text=evoBVal,  fill=white, draw=evoBVal!40,  anchor=north west, yshift=-2pt, xshift=1pt] at (axis cs:0,14.36) {\evonum{14.4}};
\node[evolbl, text=evoBHeld, fill=white, draw=evoBHeld!40, anchor=east, xshift=-2.5pt] at (axis cs:0,13.26) {\evonum{13.3}};
\node[evolbl, text=evoBXfer, fill=white, draw=evoBXfer!40, anchor=south, yshift=2.5pt] at (axis cs:0,15.81) {\evonum{15.8}};
\node[evolbl, text=evoBVal,  fill=evoBVal!3,  draw=evoBVal!40,  anchor=south, yshift=3pt]  at (axis cs:11,36.14) {\evonum{36.1}};
\node[evolbl, text=evoBHeld, fill=evoBHeld!3, draw=evoBHeld!40, anchor=east, xshift=-3pt] at (axis cs:9,39.07) {\evonum{39.1}};
\node[evolbl, text=evoBXfer, fill=evoBXfer!3, draw=evoBXfer!40, anchor=south, yshift=3pt] at (axis cs:4,27.91) {\evonum{27.9}};

\nextgroupplot[evo/pair, evo/claudex, title={(c) CC + Opus 4.6}, title style={xshift=0.05cm}]
\evoband{2}\evoband{3}
\addplot[evo/cval]  table[x=iter, y=valpair]  {\evoClaudeA};
\addplot[evo/cheld] table[x=iter, y=heldpair] {\evoClaudeA};
\addplot[evo/cxfer] table[x=iter, y=xferpair] {\evoClaudeA};
\addplot[evo/bestpair, color=evoCVal]  coordinates {(3,33.66)};
\addplot[evo/bestpair, color=evoCHeld] coordinates {(4,21.86)};
\addplot[evo/bestpair, color=evoCXfer] coordinates {(3,37.21)};
\node[evolbl, text=evoCXfer, fill=white, draw=evoCXfer!40, anchor=south, yshift=2.5pt] at (axis cs:0,22.56) {\evonum{22.6}};
\node[evolbl, text=evoCHeld, fill=white, draw=evoCHeld!40, anchor=south, yshift=2.5pt] at (axis cs:0,10.47) {\evonum{10.5}};
\node[evolbl, text=evoCVal,  fill=white, draw=evoCVal!40,  anchor=north, yshift=-2.5pt] at (axis cs:0,10.40) {\evonum{10.4}};
\node[evolbl, text=evoCXfer, fill=evoCXfer!3, draw=evoCXfer!40, anchor=south, yshift=3pt]   at (axis cs:3,37.21) {\evonum{37.2}};
\node[evolbl, text=evoCVal,  fill=evoCVal!3,  draw=evoCVal!40,  anchor=east, xshift=-3.5pt] at (axis cs:3,33.66) {\evonum{33.7}};
\node[evolbl, text=evoCHeld, fill=evoCHeld!3, draw=evoCHeld!40, anchor=south, yshift=3pt]   at (axis cs:4,21.86) {\evonum{21.9}};
\nextgroupplot[evo/pair, evo/glmx, title={(d) CC + GLM-5.3}]
\evoband{6}
\addplot[evo/dval] table[x=iter, y=valpair] {\evoClaudeB};
\addplot[evo/dheld] table[x=iter, y=heldpair] {\evoClaudeB};
\addplot[evo/dxfer] table[x=iter, y=xferpair] {\evoClaudeB};
\addplot[evo/bestpair, color=evoDVal] coordinates {(4,31.19) (6,31.19)};
\addplot[evo/bestpair, color=evoDHeld] coordinates {(1,25.58)};
\addplot[evo/bestpair, color=evoDXfer] coordinates {(8,20.70)};
\node[evolbl, text=evoDXfer, fill=white, draw=evoDXfer!40, anchor=north, yshift=-2.5pt] at (axis cs:0,14.42) {\evonum{14.4}};
\node[evolbl, text=evoDXfer, fill=evoDXfer!3, draw=evoDXfer!40, anchor=north, yshift=-3pt] at (axis cs:8,20.70) {\evonum{20.7}};
\node[evolbl, text=evoDVal, fill=white, draw=evoDVal!40, anchor=south, yshift=2.5pt] at (axis cs:0,19.80) {\evonum{19.8}};
\node[evolbl, text=evoDHeld, fill=white, draw=evoDHeld!40, anchor=north west, xshift=1pt, yshift=-1pt] at (axis cs:0,18.37) {\evonum{18.4}};
\node[evolbl, text=evoDHeld, fill=evoDHeld!3, draw=evoDHeld!40, anchor=north west, xshift=2pt, yshift=-2.5pt] at (axis cs:1,25.58) {\evonum{25.6}};
\node[evolbl, text=evoDVal, fill=evoDVal!3, draw=evoDVal!40, anchor=south, yshift=3pt] at (axis cs:5,31.19) {\evonum{31.2}};
\nextgroupplot[evo/joint, evo/codexx, evo/legend, legend style={at={(0,-0.45)}, anchor=north west}, ylabel={Joint (\%)}]
\evoband{6}\evoband{8}\evoband{12}
\addplot[evo/aval]  table[x=iter, y=valjoint]  {\evoCodexA}; \addlegendentry{Validation (Inkling-Small)}
\addplot[evo/aheld] table[x=iter, y=heldjoint] {\evoCodexA}; \addlegendentry{Held-out (Inkling-Small)}
\addplot[evo/axfer] table[x=iter, y=xferjoint] {\evoCodexA}; \addlegendentry{Transfer (Nemotron)}
\nextgroupplot[evo/joint, evo/codexx, evo/legend, legend style={xshift=-0.25cm}]
\evoband{1}\evoband{5}\evoband{9}\evoband{11}\evoband{14}
\addlegendimage{evo/bval}\addlegendentry{Validation (DeepSeek)}
\addlegendimage{evo/bheld}\addlegendentry{Held-out (DeepSeek)}
\addlegendimage{evo/bxfer}\addlegendentry{Transfer (GLM-4.7 Flash)}
\addplot[evo/noprompt, forget plot] coordinates {(1,28.2) (4,25.8)};
\addplot[evo/noprompt, forget plot] coordinates {(1,3.0) (4,7.8)};
\addplot[evo/noprompt, forget plot] coordinates {(1,19.4) (4,11.9)};
\addplot[evo/bval,  forget plot] table[x=iter, y=valjoint]  {\evoCodexB};
\addplot[evo/bheld, forget plot] table[x=iter, y=heldjoint] {\evoCodexB};
\addplot[evo/bxfer, forget plot] table[x=iter, y=xferjoint] {\evoCodexB};
\nextgroupplot[evo/joint, evo/claudex, evo/legend, legend style={at={(0,-0.45)}, anchor=north west, xshift=-0.35cm}]
\evoband{2}\evoband{3}
\addplot[evo/cval]  table[x=iter, y=valjoint]  {\evoClaudeA}; \addlegendentry{Validation (Opus 4.6)}
\addplot[evo/cheld] table[x=iter, y=heldjoint] {\evoClaudeA}; \addlegendentry{Held-out (Opus 4.6)}
\addplot[evo/cxfer] table[x=iter, y=xferjoint] {\evoClaudeA}; \addlegendentry{Transfer (Opus 5)}
\nextgroupplot[evo/joint, evo/glmx, evo/legend, legend style={at={(1,-0.45)}, anchor=north east, xshift=0.6cm}]
\evoband{6}
\addplot[evo/dval] table[x=iter, y=valjoint] {\evoClaudeB}; \addlegendentry{Validation (GLM-5.3)}
\addplot[evo/dheld] table[x=iter, y=heldjoint] {\evoClaudeB}; \addlegendentry{Held-out (GLM-5.3)}
\addplot[evo/dxfer] table[x=iter, y=xferjoint] {\evoClaudeB}; \addlegendentry{Transfer (gpt-oss-120b)}
\end{groupplot}
\node[anchor=north, yshift=-5pt, draw=gray!45, rounded corners=1.5pt, inner xsep=5pt, inner ysep=3pt,
      font=\fontsize{7.5}{8.5}\selectfont] at (current bounding box.south)
    {\tikz[baseline=-0.55ex]\fill[yellow!30] (0,-0.13) rectangle (0.42,0.13);\ Pareto front (validation)\qquad
     \tikz[baseline=-0.55ex]\draw[black, densely dotted, line width=0.5pt] (0,0) -- (0.42,0);\ No metric};
\end{tikzpicture}

\caption[Pairwise accuracy and joint hit rate across discovery iterations]{\textbf{Pairwise accuracy and joint hit rate across discovery iterations.} Pairwise accuracy (top) and joint hit rate (bottom) across optimizer iterations for four optimization sessions. CC = Claude Code; DS = DeepSeek. Validation uses TitanVul+CWEval; held-out and cross-model transfer use PrimeVul. \capstar{} marks each series' best pairwise accuracy, \capband{} marks the validation Pareto front, and labels show seed and best values. Harnesses are selected using validation only; PrimeVul is never used for selection. No metric marks iterations where the optimizer failed or returned an invalid harness.}
\label{fig:pairwise-evolution}

\end{figure}
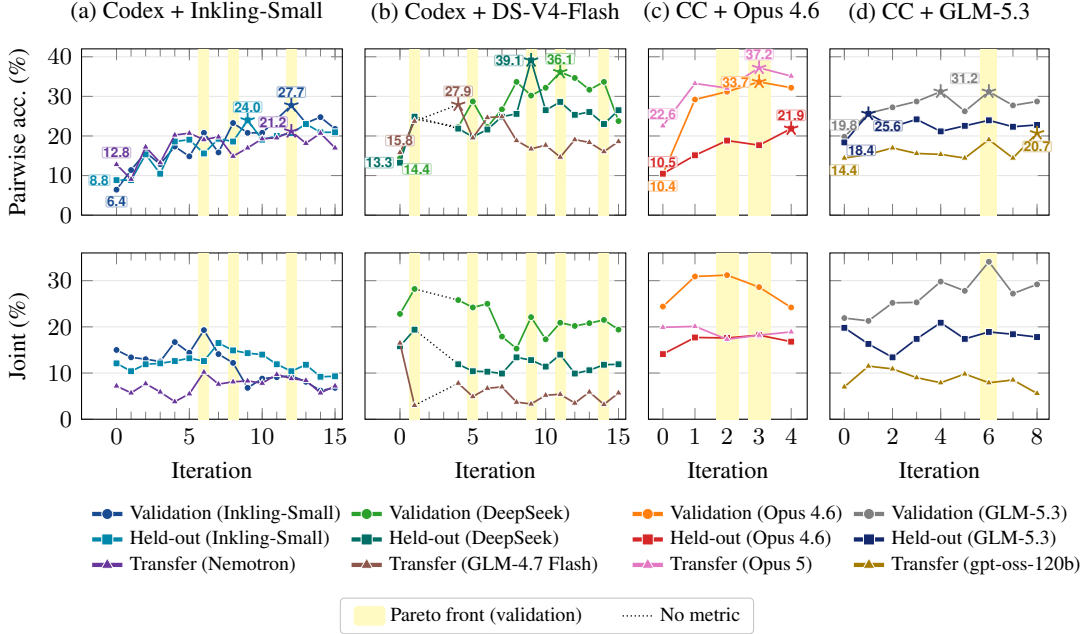

%% file: results/related_comparison.tex
\begin{table}[t]
\centering
\caption[GPT-4o-based methods on PrimeVul]{\textbf{GPT-4o-based methods on PrimeVul.} P-C is the number of pairs for which both the vulnerable and patched functions are classified correctly; Error is the number of pairs in which at least one of the two functions is misclassified; and FPR is the false-positive rate. $\uparrow$ means higher is better, $\downarrow$ means lower is better, and bold marks the best result in each primary metric.}
\label{tab:primevul-llm-performance}
\setlength{\tabcolsep}{4.5pt}
\renewcommand{\arraystretch}{1.08}
\resizebox{\textwidth}{!}{%
\begin{tabular}{lrrrrrrr}
\toprule
\textbf{Method}
& \textbf{P-C} $\uparrow$
& \textbf{Error} $\downarrow$
& \makecell{\textbf{Precision} $\uparrow$}
& \makecell{\textbf{Recall} $\uparrow$}
& \textbf{FPR} $\downarrow$
& \makecell{\textbf{Accuracy} $\uparrow$}
& \makecell{\textbf{F1 score} $\uparrow$} \\
\midrule
Ding et al.'s CoT~\citep{primevul}
& 40 & 395
& 0.5355 & 0.1908 & 0.1655 & 0.5126 & 0.2814 \\
GPTLens~\citep{gptlens}
& 44 & 391
& 0.5144 & 0.6552 & 0.6184 & 0.5184 & 0.5763 \\
VulTrial~\citep{vultrial}
& 81 & 354
& 0.5317 & 0.5977 & 0.5264 & 0.5356 & 0.5628 \\
ConfColl~\citep{confcoll}
& 22 & 413
& 0.5078 & 0.8253 & 0.8000 & 0.5126 & \textbf{0.6287} \\
\midrule
Our approach (seed)
& 74 & 361
& 0.5185 & 0.5793 & 0.5379 & 0.5207 & 0.5472 \\
Our approach (iter 12 harness)
& \textbf{103} & \textbf{332}
& 0.5636 & 0.4276 & 0.3310 & \textbf{0.5483} & 0.4863 \\
\midrule
\multicolumn{8}{l}{\textit{Our approach on 430 pairs, validation overlap removed$^{\dagger}$}} \\
Our approach (seed)
& 71 & 359
& 0.5157 & 0.5744 & 0.5395 & 0.5174 & 0.5435 \\
Our approach (iter 12 harness)
& 100 & 330
& 0.5631 & 0.4256 & 0.3302 & 0.5477 & 0.4848 \\
\bottomrule
\end{tabular}%
}

\begin{minipage}{\textwidth}
\footnotesize
\textit{Notes.}
Bold marks the best result among methods evaluated on all 435 pairs. $^{\dagger}$Excludes 5 pairs overlapping validation.
\end{minipage}
\end{table}

%% file: results/shortcut_table.tex
\begin{table}[t]
\centering
\small
\caption{\textbf{Shortcut checks for Pareto-frontier harnesses.} The ratings indicate how many of the frontier harnesses receive a $4/4$ score for preserving SCoT, avoiding prohibited content, and preserving the output format. The final columns show how much the frontier harnesses cut FP (false positives) and total flagged ``vulnerable'' functions, compared to the run's seed. \supp{S2_harness_quality}{S2} shows more detailed results.}
\label{tab:shortcut-main}
\resizebox{\linewidth}{!}{%
\begin{tabular}{llrrrrr}
\toprule
\textbf{Agent + Model} & \textbf{Frontier iterations} & \textbf{SCoT} & \textbf{No prohibited} & \textbf{Output} & \textbf{FP reduction} & \textbf{Flagged reduction} \\
\midrule
Codex + Inkling-Small      & 6, 8, 12        & 3/3 & 3/3 & 3/3 & 33-44\% & 31-36\% \\
Codex + DeepSeek V4 Flash  & 1, 5, 9, 11, 14 & 5/5 & 5/5 & 4/5 & 44-60\% & 34-49\% \\
Claude Code + Opus 4.6     & 2, 3            & 2/2 & 2/2 & 2/2 & 50\%     & 35-37\% \\
Claude Code + GLM-5.3      & 6               & 1/1 & 1/1 & 0/1 & 49\%     & 35\% \\
\midrule
\textbf{Total} & \textbf{11 harnesses} & \textbf{11/11} & \textbf{11/11} & \textbf{9/11} & \textbf{33-60\%} & \textbf{31-49\%} \\
\bottomrule
\end{tabular}%
}
\end{table}